\documentclass[twoside,twocolumn,9pt]{article}
\usepackage{extsizes}
\usepackage[super,sort&compress,comma]{natbib} 
\usepackage[version=3]{mhchem}
\usepackage[left=1.5cm, right=1.5cm, top=1.785cm, bottom=2.0cm]{geometry}
\usepackage{balance}
\usepackage{mathptmx}
\usepackage{sectsty}
\usepackage{graphicx} 
\usepackage{lastpage}
\usepackage[format=plain,justification=justified,singlelinecheck=false,font={stretch=1.125,small,sf},labelfont=bf,labelsep=space]{caption}
\usepackage{float}
\usepackage{fancyhdr}
\usepackage{fnpos}
\usepackage[english]{babel}
\addto{\captionsenglish}{%
  
}
\usepackage{array}
\usepackage{droidsans}
\usepackage{charter}
\usepackage[T1]{fontenc}
\usepackage[usenames,dvipsnames]{xcolor}
\usepackage{setspace}
\usepackage[compact]{titlesec}
\usepackage{hyperref}

\usepackage{array,multirow}
\usepackage{tabularx}
\makeatletter
\def\hlinewd#1{%
\noalign{\ifnum0=`}\fi\hrule \@height #1 %
\futurelet\reserved@a\@xhline}

\usepackage{epstopdf}

\definecolor{cream}{RGB}{222,217,201}

\begin{document}

\pagestyle{fancy}
\thispagestyle{plain}
\fancypagestyle{plain}{
\renewcommand{\headrulewidth}{0pt}
}

\makeFNbottom
\makeatletter
\renewcommand\LARGE{\@setfontsize\LARGE{15pt}{17}}
\renewcommand\Large{\@setfontsize\Large{12pt}{14}}
\renewcommand\large{\@setfontsize\large{10pt}{12}}
\renewcommand\footnotesize{\@setfontsize\footnotesize{7pt}{10}}
\makeatother

\renewcommand{\thefootnote}{\fnsymbol{footnote}}
\renewcommand\footnoterule{\vspace*{1pt}%
\color{cream}\hrule width 3.5in height 0.4pt \color{black}\vspace*{5pt}} 
\setcounter{secnumdepth}{5}

\makeatletter 
\renewcommand\@biblabel[1]{#1}            
\renewcommand\@makefntext[1]%
{\noindent\makebox[0pt][r]{\@thefnmark\,}#1}
\makeatother 
\renewcommand{\figurename}{\small{Fig.}~}
\sectionfont{\sffamily\Large}
\subsectionfont{\normalsize}
\subsubsectionfont{\bf}
\setstretch{1.125} 
\setlength{\skip\footins}{0.8cm}
\setlength{\footnotesep}{0.25cm}
\setlength{\jot}{10pt}
\titlespacing*{\section}{0pt}{4pt}{4pt}
\titlespacing*{\subsection}{0pt}{15pt}{1pt}


\makeatletter 
\newlength{\figrulesep} 
\setlength{\figrulesep}{0.5\textfloatsep} 

\newcommand{\topfigrule}{\vspace*{-1pt}%
\noindent{\color{cream}\rule[-\figrulesep]{\columnwidth}{1.5pt}} }

\newcommand{\botfigrule}{\vspace*{-2pt}%
\noindent{\color{cream}\rule[\figrulesep]{\columnwidth}{1.5pt}} }

\newcommand{\dblfigrule}{\vspace*{-1pt}%
\noindent{\color{cream}\rule[-\figrulesep]{\textwidth}{1.5pt}} }

\makeatother

\twocolumn[
  \begin{@twocolumnfalse}
\par
\vspace{1em}
\sffamily
\begin{tabular}{m{4.5cm} p{13.5cm} }

& \noindent\LARGE{\textbf{Theoretical study of the stability of the tetradymite-like phases of Sb$_2$S$_3$, Bi$_2$S$_3$, and Sb$_2$Se$_3$}} \\ 

\vspace{0.3cm} & \vspace{0.3cm} \\
 & \noindent\large{E. Lora da Silva,$^{\ast}$\textit{$^{a}$} J. M. Skelton,\textit{$^{b}$} P. Rodr\'{i}guez-Hern\'{a}ndez, \textit{$^{c}$} A. Mu\~{n}oz, \textit{$^{c}$}  M. C. Santos, \textit{$^{d,e}$} D. Mart\'{i}nez-Garc\'{i}a, \textit{$^{f}$} R. Vilaplana, \textit{$^{g}$} and F. J. Manj\'{o}n \textit{$^{e}$}} \\ 

 & \noindent\normalsize{
    We report a comparative theoretical study of the \textit{Pnma} and \textit{R-3m} phases of Sb$_2$S$_3$, Bi$_2$S$_3$, and Sb$_2$Se$_3$ close to ambient pressure.
    Our enthalpy calculations at 0 K show that at ambient pressure the \textit{R-3m} (tetradymite-like) phase of Sb$_2$Se$_3$ is energetically more stable than the \textit{Pnma} phase, contrary to what is observed for Sb$_2$S$_3$ and Bi$_2$S$_3$, and irrespective of the exchange-correlation functional employed in the calculations.
    The result for Sb$_2$Se$_3$ is in contradiction to experiments where all three compounds are usually grown in the \textit{Pnma} phase.
    This result is further confirmed by free-energy calculations taking into account the temperature dependence of the unit-cell volumes and phonon frequencies.
    Lattice dynamics and elastic tensor calculations further show that both \textit{Pnma} and \textit{R-3m} phases of Sb$_2$Se$_3$ are dynamically and mechanically stable at zero applied pressure.
    Since these results suggest that the formation of the \textit{R-3m} phase for Sb$_2$Se$_3$ should be feasible at close to ambient conditions, we provide a theoretical crystal structure and simulated Raman and infrared spectra to help in its identification. We also discuss the results of the two published works that have claimed to have synthesized tetradymite-like Sb$_2$Se$_3$.
    Finally, the stability of the \textit{R-3m} phase across the three group-15 \textit{A$_2$X$_3$} sesquichalcogenides is analysed based on their van der Waals gap and X-X in-plane geometry.
    } \\

\end{tabular}

\end{@twocolumnfalse} \vspace{0.6cm}

  ]

\renewcommand*\rmdefault{bch}\normalfont\upshape
\rmfamily
\section*{}
\vspace{-1cm}


\footnotetext{\textit{$^{a}$~IFIMUP, Departamento de Física e Astronomia, Faculdade de Ci\^{e}nicas da Universidade do Porto, 4169-007 Porto, Portugal. Fax: +351 22 04 02 406; Tel: +351 22 04 02 362; E-mail: estelina.silva@fc.up.pt}}
\footnotetext{\textit{$^{b}$~Department of Chemistry, University of Manchester, Oxford Road, Manchester M13 9PL, United Kingdom}}
\footnotetext{\textit{$^{c}$~Departamento de F\'{i}sica, Instituto de Materiales y Nanotecnología, MALTA Consolider Team, Universidad de La Laguna, 38206 Tenerife, Spain}}
\footnotetext{\textit{$^{d}$~Sede do Agrupamento Escolas de Ponte de Sor, 7400-259 Ponte de Sor, Portugal }}
\footnotetext{\textit{$^{e}$~Instituto de Dise\~{n}o para la Fabricaci\'{o}n y Producci\'{o}n Automatizada, MALTA Consolider Team, Universitat Polit\`{e}cnica de Val\`{e}ncia, 46022 Val\`{e}ncia, Spain}}
\footnotetext{\textit{$^{f}$~Departamento de F\'{i}sica Aplicada - ICMUV, MALTA Consolider Team, Universitat de Val\`{e}ncia, 46100 Burjassot, Spain}}
\footnotetext{\textit{$^{g}$~Centro de Tecnologías F\'{i}sicas, MALTA Consolider Team, Universitat Polit\`{e}cnica de Val\`{e}ncia, 46022 Val\`{e}ncia, Spain}}




\section{Introduction}
 
\begin{figure*}[h]
\begin{center}
\includegraphics[width=14cm]{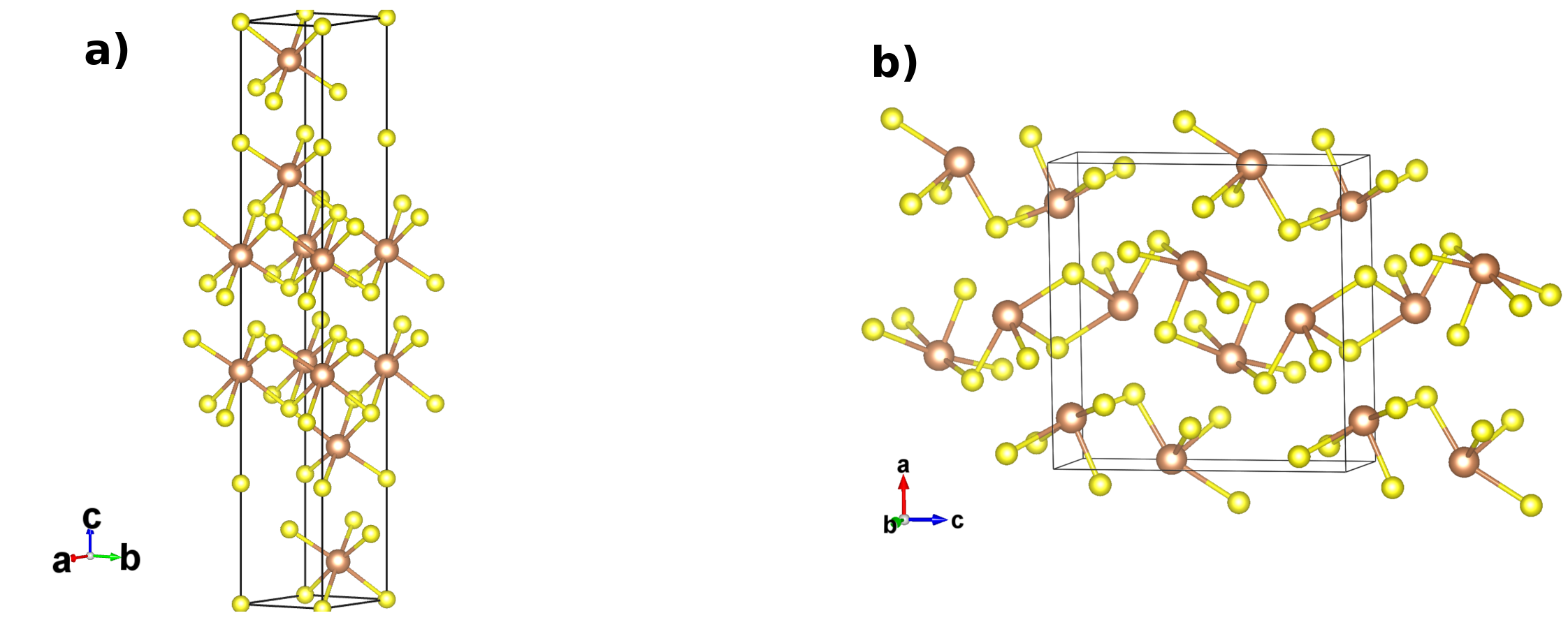}
\caption{\label{fig:unit-cell}
Images of the \textit{R-3m} (a) and \textit{Pnma} (b) phases of the \textit{A$_2$X$_3$} sesquichalcogenides  ($A$ = Sb, Bi; $X$ = S, Se). The $A$ cations and $X$ anions are shown as brown and yellow spheres, respectively.} 
\end{center}
\end{figure*}

Since the identification of the trigonal tetradymite-like \textit{R-3m} phases of group-15 sesquichalcogenides (i.e. Sb$_2$Te$_3$, Bi$_2$Se$_3$, Bi$_2$Te$_3$; \ref{fig:unit-cell}) as 3D topological insulators,\cite{Science.325.178.2009, NatPhys.5.438.2009} the family of \textit{A$_2$X$_3$} sesquichalcogenides has attracted a great deal of attention from the scientific community.
Three-dimensional topological insulators represent a new class of matter, with insulating bulk electronic states and topologically-protected metallic surface states arising from time-reversal symmetry and strong spin-orbit coupling.
These properties make them of potential interest for spintronics and quantum computing applications.\cite{RevModPhys.282.3045.2010}
Due to this fundamental interest and potential applications, the identification of new topological insulators and materials with superconducting properties is currently an important research area in condensed matter science.

The stibnite (Sb$_2$S$_3$), bismuthinite (Bi$_2$S$_3$), and antimonselite (Sb$_2$Se$_3$) minerals are also group-15 sesquichalcogenides but do not crystallize in the tetradymite-like \textit{R-3m} structure under ambient conditions but instead adopt the orthorhombic U$_2$S$_3$-type \textit{Pnma} structure (Fig. \ref{fig:unit-cell}.b).
Sb$_2$S$_3$, Bi$_2$S$_3$, and Sb$_2$Se$_3$ are semiconductors with band-gap widths of 1.7, 1.3, and 1.2 eV respectively.\cite{JPhysChemSolids.2.240.1957,PhysRevB.87.205125.2013}
These materials are used in a wide range of technological applications including photovoltaics (solar cells), X-ray computed tomography detectors, fuel cells, gas sensors, and for detection of biomolecules.\cite{JPhysChemLett.1.1524.2010,AdvFunctMater.21.4663.2011,NatPhotonics.9.409.2015,NatEnergy.2.17046.2017,NatMater.5.118.2006,PhysChemB.110.21408.2006,NanoLett.9.1482.2009}
Additionally, Sb$_2$Se$_3$ has recently found a number of other applications including in solid-state batteries, fiber lasers, and photoelectrochemical devices.\cite{SolEnergMat.230.111223.2021,CHEN2022508,AdvFunctMater.2112776.2022,MA2021107286}

Since several phases, including the \textit{Pnma} phase, have been synthesized for Bi$_2$Se$_3$, which usually crystallizes in the tetradymite-like \textit{R-3m} structure,\cite{KANG2017223,Kulbachinskii_2018,acs.jpcc.6b02559,pssb.257.202000145.2020} it is natural to wonder whether the \textit{R-3m} structure could be adopted by other sesquichalcogenides that generally adopt the U$_2$S$_3$-type structure, \textit{viz.} Sb$_2$S$_3$, Bi$_2$S$_3$, and in particular Sb$_2$Se$_3$.
In fact, several theoretical studies have been performed over the years to investigate the properties of the hypothetical tetradymite-like Sb$_2$Se$_3$ structure.
Some of these works have suggested that this phase should undergo a topological quantum phase transition under compression,\cite{PhysRevB.84.245105.2011,PhysRevB.89.035101.2014} while one found that tetradymite-like Sb$_2$Se$_3$ is dynamically stable and is a topological insulator at ambient pressure.\cite{PhysRevB.97.075147.2018}
Interestingly enough, the \textit{R-3m} phase of group-15 sesquichalcogenides is characterized by a unique type of bonding termed "metavalent bonding" that not only underpins the topological properties but also makes these materials useful for phase-change memories, as highly efficient thermoelectrics, and for photonic devices.\cite{adma.201908302}

In 2013 an experimental study claimed to have observed the \textit{R-3m} phase of Sb$_2$Se$_3$, with a topological transition occurring at 2 GPa,\cite{PhysRevLett.110.107401.2013} but the tetradymite-like structure of Sb$_2$Se$_3$ was not confirmed beyond doubt.
On the other hand, a comparative experimental and theoretical study of the three U$_2$S$_3$-type sesquichalcogenides suggested the \textit{Pnma} structure to be stable up to 50 GPa.\cite{JPhysChemC.120.10547.2016}
This is supported by several experimental high-pressure studies on Sb$_2$Se$_3$ in which the \textit{Pnma} structure was found to be stable up to 50 GPa and above.\cite{SciRep.3.2665.2013,ApplPhysLett.108.212601.2016}
However, one study observed a pressure-induced isostructural phase transition at 12 GPa and a further transition to a disordered \textit{Im-3m} structure above 50 GPa, followed by a pressure-induced amorphization on releasing the pressure.\cite{acs.jpcc.9b11520}
Despite the apparent stability of the \textit{Pnma} phase, experimental high-pressure studies have also found that \textit{Pnma}-type Sb$_2$Se$_3$ becomes a topological superconductor at 2.5 K and around 10 GPa,\cite{SciRep.4.6679.2014} exhibiting highly-conducting spin-polarized surface states similar to those observed for Bi$_2$Se$_3$.\cite{PhysRevB.97.235306.2018}
Furthermore, a recent study has claimed to have synthesized the \textit{R-3m} structure of Sb$_2$Se$_3$  by atomic layer epitaxy on a buffer layer of Bi$_2$Se$_3$.\cite{Matetskiy_2020}
We can therefore conclude that while the bulk of the experimental evidence suggests the \textit{R-3m} phase of Sb$_2$Se$_3$ is not observed at high pressure, it is not conclusive whether this phase could potentially be formed under favourable synthesis conditions.

In the light of the above studies, it is interesting to compare the stabilities of the \textit{Pnma} and \textit{R-3m} structural phases of the the three U$_2$S$_3$-type sesquichalcogenides at close to ambient conditions, and to confirm whether or not the \textit{R-3m} phase could be synthetically accessible.
In this work, we report a set of systematic density-functional theory (DFT) calculations on the \textit{Pnma} and \textit{R-3m} phases of the three U$_2$S$_3$-type sesquichalcogenides under ambient conditions and at pressures of up to 10 GPa.
We show that the \textit{Pnma} phase of Sb$_2$S$_3$ and Bi$_2$S$_3$ is energetically more stable than the \textit{R-3m} phase over this pressure range, but that, unexpectedly, the \textit{R-3m} phase of Sb$_2$Se$_3$ is predicted by several exchange-correlation (xc) functionals to be more stable than the \textit{Pnma} phase close to ambient conditions. 
To aid in future experimental efforts to prepare the \textit{R-3m} phase of Sb$_2$Se$_3$, we also confirm its dynamical and mechanical stability and provide a theoretical structure and vibrational spectra to support its identification.
Finally, we also discuss the only two works that, to our knowledge, have claimed to have prepared the tetradymite-like phase of Sb$_2$Se$_3$ at close to ambient conditions.

\section{Methods}

The structural properties of the different crystalline phases of Sb$_2$S$_3$, Bi$_2$S$_3$, and Sb$_2$Se$_3$ were calculated within the framework of pseudopotential plane-wave density-functional theory.\cite{hohenberg-pr-136-1964}
The Vienna \textit{Ab-initio} Simulation Package (VASP) code \cite{kresse-cms-6-1996} was employed to perform simulations.
The revised Perdew-Burke-Ernzerhof generalized-gradient approximation (GGA) functional for solids (PBEsol) \cite{perdew-prl-100-2008,perdew-prl-102-2009} was used for all calculations.
Additional calculations were also performed with the local-density approximation (LDA)\cite{PhysRevB.23.5048} functional and the dispersion-corrected PBE-D2 \cite{JCompChem.27.1787.2006} functional to examine the effect of the xc functional on the results.
Projector augmented-wave (PAW) pseudopotentials including six valence electrons for S[3s$^2$3p$^4$] and Se[4s$^2$4p$^4$] and fifteen valence electrons for Sb[4d$^{10}$5s$^2$5p$^3$] and Bi[5d$^{10}$6s$^2$6p$^3$] were used to model the ion cores.
Convergence of the total energy was achieved with a plane-wave kinetic-energy cut-off of 600 eV.
The Brillouin-zone (BZ) was sampled with $\Gamma$-centered Monkhorst-Pack \cite{monkhorst-prb-13-1976} grids with appropriate subdivisions for the different structural phases of the three compounds, \textit{viz.}: \textit{Pnma} - 6 $\times$ 10 $\times$ 6 and \textit{R-3m} - 12 $\times$ 12 $\times$ 12.

Structural relaxations were performed by allowing the atomic positions and the unit-cell parameters to optimise at a series of different volumes in order to confirm the stability of both the \textit{Pnma} and \textit{R-3m} phases in a pressure range from 0-10 GPa. 
At each volume we obtain the (isotropic) external pressure for the applied compression
and the corresponding structural parameters.
The pressure-volume ($p$-$V$) curves for each of the compounds were fitted to a third-order Birch-Murnaghan equation of state \cite{murnaghan-pnas-30-1944,birch-pr-71-1947} to obtain the equilibrium volume, the bulk modulus, and its pressure derivative.
The enthalpy $H$ as a function of volume were computed using the relation $H=E+pV$, where $E$ is the total electronic energy of the system, $p$ is pressure, and $V$ is the volume.
Comparison of the $H$ curves of the different polymorphs can provide insight into the relative thermodynamic stabilities over the studied pressure range.

Lattice-dynamics calculations were performed on the \textit{Pnma} and \textit{R-3m} phases of Sb$_2$Se$_3$ at a series of cell volumes corresponding to different applied pressures.
The phonon frequencies were computed by using the supercell finite-displacement method implemented in the Phonopy package \cite{togo-prb-78-2008} with VASP as the force calculator. \cite{chaput-prb-84-2001}
Supercell expansions of 2 $\times$ 4 $\times$ 2 for the \textit{Pnma} phase and 2 $\times$ 2 $\times$ 2 for the \textit{R-3m} phases were used to enable the exact calculation of frequencies at the zone center ($\Gamma$) and unique zone-boundary wavevectors, which were interpolated to obtain phonon-dispersion curves together with density of states curves on uniform 50 $\times$ 50 $\times$ 50 $\Gamma$-centered \textbf{q}-point meshes.

Infrared (IR) and Raman spectra were calculated for the ground-state \textit{R-3m} phase of Sb$_2$Se$_3$ using the methods described in Ref. \cite{PhysChemChemPhys.19.12452.2017} and implemented in the Phonopy-Spectroscopy package.\cite{PhonoptSpec}
The spectral linewidths were obtained by computing the third-order force constants, of a 2 $\times$ 2 $\times$ 2 expansion of the primitive-cell, and following the many-body perturbative approach described in detail in Refs. \cite{PhysChemChemPhys.19.12452.2017} and \cite{PhysRevB.91.094306.2015} and implemented in the Phono3py software.\cite{PhysRevB.91.094306.2015}

Elastic tensors were computed to assess the mechanical stability of the \textit{Pnma} and \textit{R-3m} phases of Sb$_2$Se$_3$ at zero pressure, by employing the central-difference method where the unique components of the elastic tensor are determined by performing six finite distortions of the lattice and deriving the tensor elements from the strain-stress relationship.\cite{PhysRevB.65.104104.2002}
For these calculations, it was necessary to increase the plane-wave energy cutoff to converge the stress tensor, which was achieved by systematically increasing the plane-wave cutoff up to 950 eV.
We then employed the ELATE software \cite{JPhysCondensMatter.28.275201.2016} to analyze the linear compressibility using the computed stress tensors.

\section{Results and Discussion}
\subsection{Structural properties of the \textit{Pnma} phase}

In order to verify the accuracy of our theoretical calculations as a prior step before attempting to study of the potential \textit{R-3m} phases of Sb$_2$S$_3$, Bi$_2$S$_3$ and Sb$_2$Se$_3$, we first calculated the equilibrium lattice parameters, bulk moduli and pressure derivatives of the \textit{Pnma} phases and compared them to other experimental and theoretical studies in the literature (Tab. \ref{table:pnma_param}).

\renewcommand{\thefootnote}{\alph{footnote}}

\begin{table*}
\begin{center}
\small
\caption{ \label{table:pnma_param} 
Calculated equilibrium lattice parameters (a$_0$, b$_0$ and c$_0$), equilibrium bulk moduli (B$_0$) and pressure derivatives (B$_0'$) of the \textit{Pnma} phases of Sb$_2$Se$_3$, Sb$_2$S$_3$ and Bi$_2$S$_3$. Values are compared to experiments and other theoretical results from the literature.}
\begin{tabular}{|c |c|c|c|c|c|c |}\hlinewd{1pt}
 & \multicolumn{2}{c|}{\textbf{Sb$_2$Se$_3$}}  & \multicolumn{2}{c|}{\textbf{Sb$_2$S$_3$}} & \multicolumn{2}{c|}{\textbf{Bi$_2$S$_3$ }}   \\ \hline
 \multirow{5}{*}{\textbf{a$_0$ (\AA)}} & \multicolumn{2}{c|}{11.75$^a$}  & \multicolumn{2}{c|}{11.24$^a$} & \multicolumn{2}{c|}{11.19$^a$}   \\ \cline{2-7}
 &  \textit{Theo.} &	\textit{Exp.} & \textit{Theo.} &	\textit{Exp.} & \textit{Theo.} &	\textit{Exp.}\\\cline{2-7}
 & 11.80$^b$ &  11.80$^{f}$  &  11.27$^b$  	&   11.30$^{b,j,k}$  &  11.41$^b$  &  11.27$^p$ \\
 & 11.52$^c$  &   11.79$^g$   &  11.02$^c$  	&    11.31$^{l,m}$   &  11.00$^{n}$  & 11.33$^q$ \\
 & 11.91$^d$ &				&	11.30$^h$	&				     &	11.58$^{o}$	&			\\
 & 11.53$^e$ &				 &	11.08$^i$	&				     &				&			\\\hline
 \multirow{5}{*}{\textbf{b$_0$ (\AA)}} & \multicolumn{2}{c|}{3.98$^a$}  & \multicolumn{2}{c|}{3.83$^a$} & \multicolumn{2}{c|}{3.96$^a$}   \\ \cline{2-7}
 &   \textit{Theo.} &	\textit{Exp.} & \textit{Theo.} &	\textit{Exp.} & \textit{Theo.} &	\textit{Exp.}\\\cline{2-7}
  & 3.99$^{b}$  &  3.98$^{f}$  &  3.81$^c$  & 3.84$^{b,j,k}$  &  3.97$^b$  &  3.97$^p$ \\
  & 3.96$^{c,e}$ & 3.99$^{g}$ &  3.84$^h$  & 	3.84$^{l,m}$ 	&  3.94$^{n}$  & 3.98$^q$ 		 \\
  & 3.98$^d$   &			   &	3.83$^{b,i}$	   &				    &	3.99$^{o}$	&    	\\\hline
  \multirow{5}{*}{\textbf{c$_0$ (\AA)}} & \multicolumn{2}{c|}{11.30$^a$}  & \multicolumn{2}{c|}{10.91$^a$} & \multicolumn{2}{c|}{10.94$^a$}   \\ \cline{2-7}
 &   \textit{Theo.} &	\textit{Exp.} & \textit{Theo.} &	\textit{Exp.} & \textit{Theo.} &	\textit{Exp.}\\\cline{2-7}
  & 11.28$^b$   &  11.65$^{f,g}$  &  10.89$^b$  & 11.23$^{b,j,l,m}$  &  11.01$^b$  &  11.13$^p$ \\
  & 11.22$^{c,e}$  &  	 	    &  10.79$^c$  &  11.24$^{k}$ 		& 	10.83$^{n}$	  & 11.18$^q$  \\
  & 11.70$^d$ 	&				&	11.22$^h$&				       &	11.05$^{o}$	&\\
  &				&				&	10.81$^i$&				      &				& \\ \hline\hline
     \multirow{5}{*}{\textbf{V$_0$ (\AA$^3$)}} & \multicolumn{2}{c|}{528.11$^a$}  & \multicolumn{2}{c|}{469.6$^a$} & \multicolumn{2}{c|}{484.4$^a$}   \\ \cline{2-7}
 &   \textit{Theo.} &	\textit{Exp.} & \textit{Theo.} &	\textit{Exp.} & \textit{Theo.} &	\textit{Exp.}\\\cline{2-7}
  & 531.1$^b$  &  547.1$^{f}$    &  470.4$^b$  & 486.0$^{b}$      &  498.3$^b$  &  498.4$^p$ \\
  &  511.8$^c$  &  547.5$^g$    &   453.0$^c$  & 487.7$^{i,m,j}$    & 469.1$^{n}$ &  501.6$^q$ \\
  &  598.1$^r$  & 552.5$^s$    	&   529.9$^r$  & 488.2$^{k}$      & 510.1$^{o}$    &            \\
    &  			 & 			  	&   			 & 				    & 511.6$^q$    &            \\
\hline
    \multirow{5}{*}{\textbf{B$_0$ (GPa)}} & \multicolumn{2}{c|}{31.1$^a$}  & \multicolumn{2}{c|}{31.5$^a$} & \multicolumn{2}{c|}{42.3$^a$}   \\ \cline{2-7}
 &   \textit{Theo.} &	\textit{Exp.} & \textit{Theo.} &	\textit{Exp.} & \textit{Theo.} &	\textit{Exp.}\\\cline{2-7}
  &  70.5$^c$ &  30.0$^{f}$  &  32.2$^b$  & 37.6$^{b}$   &  	83.6$^{n}$   &  36.6$^p$	 \\
  &          &   32.7$^{s}$ &    80.3$^c$  &  26.9$^{j}$ & 	32.3$^{o}$  &  38.9$^q$	 \\
  &  		&				&			& 27.2$^k$		&	 36.5$^q$   	& 37.5$^u$ 	\\
    &  		&				&			& 41.4$^t$		&		     	&			\\\hline
     \multirow{5}{*}{\textbf{B$_0'$ }} & \multicolumn{2}{c|}{6.6$^a$}     & \multicolumn{2}{c|}{6.6$^a$} & \multicolumn{2}{c|}{6.8$^a$}   \\ \cline{2-7}
 &   \textit{Theo.} &	\textit{Exp.} & \textit{Theo.} &	\textit{Exp.} & \textit{Theo.} &	\textit{Exp.}\\\cline{2-7}
  &           &  6.1$^{f}$  &  6.2$^b$  & 3.8$^{b}$    & 5.9$^q$  		&  6.4$^p$  \\
  &          &  5.6$^{s}$ &  		  & 7.9$^{j}$      &  6.4$^{o}$ 	&	5.5$^q$	\\
    &         & 			&  	       & 6.0$^{k}$     &  			 &	4.6$^u$	\\
    &  		&				&		& 7.8$^t$	  &	       		   &		\\\hline
\hlinewd{1pt}  
\end{tabular}\\
\footnotemark[1]{This work,} 
\footnotemark[2]{Ref. \cite{JPhysChemC.120.10547.2016},} 
\footnotemark[3]{Ref. \cite{SolidStateSci.14.1211.2012},}  
\footnotemark[4]{Ref. \cite{EJChem.6.S147.2009},} 
\footnotemark[5]{Ref. \cite{ChemSci.6.5255.2015},} 
\footnotemark[6]{Ref. \cite{SciRep.3.2665.2013},}  
\footnotemark[7]{Ref. \cite{ZKristall.171.261.1985},}  
\footnotemark[8]{Ref. \cite{PhysicaB.406.287.2011},}  
\footnotemark[9]{Ref. \cite{PhysChemChemPhys.16.345.2014},}  
\footnotemark[10]{Ref. \cite{PhysChemMinerals.30.463.2003},}  
\footnotemark[11]{Ref. \cite{SciRep.6.24246.2016},}  
\footnotemark[12]{Ref. \cite{ZKristall.135.308.1972},}  
\footnotemark[13]{Ref. \cite{PhysChemMiner.135.29.254.2002}, } 
\footnotemark[14]{Ref. \cite{JMolModel.20.2180.2014},}  
\footnotemark[15]{Ref. \cite{CompMatSci.101.301.2015},} 
\footnotemark[16]{Ref. \cite{PhysChemMinerals.32.578.2005},}  
\footnotemark[17]{Ref. \cite{JPhysChemA.118.1713.2014},}  
\footnotemark[18]{Ref. \cite{JSolidStateChem.213.116.2014},}  
\footnotemark[19]{Ref. \cite{SciRep.4.6679.2014},}  
\footnotemark[20]{Ref. \cite{SciRep.8.14795.2018},}  
\footnotemark[21]{Ref. \cite{JAlloysCompd.688.329.2016}}  
\end{center}
\end{table*}

The \textit{Pnma} phase of the \textit{A$_2$X$_3$} sesquichalcogenides comprises layers stacked by weak interactions along the $a$-axis, the description of which is challenging for conventional DFT functionals.\cite{JSolidStateChem.213.116.2014,ChemSci.6.5255.2015,JPhysChemC.120.10547.2016}

The calculated lattice parameters of Sb$_2$Se$_3$ (a$_0$ = 11.75~\AA, b$_0$ = 3.98~\AA~ and c$_0$ = 11.30~\AA) are in good agreement with the experimental measurements in Refs. \cite{SciRep.3.2665.2013} and \cite{ZKristall.171.261.1985} (a$_0$ = 11.80~\AA, b$_0$ = 3.97~\AA, c$_0$ = 11.65~\AA~ and  a$_0$=11.79~\AA, b$_0$ = 33.98~\AA~ and c$_0$ = 11.65~\AA, respectively), and also with other \textit{ab initio} calculations. \cite{JPhysChemC.120.10547.2016, SolidStateSci.14.1211.2012, EJChem.6.S147.2009,ChemSci.6.5255.2015}
The most notable deviation of our calculated values from experimental measurements is a $\sim$3\% reduction of the c$_0$ parameter, which unsurprisingly leads to an underestimation of the V$_0$ of $\sim$3-4\% compared to experiment.
Our results are comparable to the theoretical results in Ref. \cite{JPhysChemC.120.10547.2016}, where calculations were also carried out using PAW pseudopotentials and the PBEsol functional.
The c$_0$ of 11.70~\AA~quoted in Ref. \cite{EJChem.6.S147.2009} is much larger than the present results but closer to experiments, while the a$_0$ parameter has a larger error compared to experiments.
We attribute this to the use of the PBE functional in this study, which has a tendency to overestimate volumes and has been shown to do so by $\sim$10\% for the antimony chalcogenides. \cite{JSolidStateChem.213.116.2014}).
On the other hand, the LDA tends to underestimate volumes, as can be seen in the lattice parameters quoted in Refs. \cite{ChemSci.6.5255.2015}, \cite{SolidStateSci.14.1211.2012} and \cite{JMolModel.20.2180.2014}.
Interestingly, the difference in the $b$-axis lengths between the theoretical and experimental studies are very small, which we attribute to the fact that this crystallographic direction corresponds to covalently-bonded chains of atoms.

Our calculated lattice parameters for the \textit{Pnma} phase of Sb$_2$S$_3$ show similar trends to those for Sb$_2$Se$_3$.
As shown on Tab. \ref{table:pnma_param}, the calculated parameters agree well with experimental measurements \cite{JPhysChemC.120.10547.2016, PhysChemMinerals.30.463.2003} and other theoretical results.\cite{JPhysChemC.120.10547.2016,SolidStateSci.14.1211.2012,PhysicaB.406.287.2011,PhysChemChemPhys.16.345.2014}
We note, however, and as for Sb$_2$Se$_3$ the lattice parameters obtained from LDA calculations tend to underestimate compared to experiments, resulting in discrepancies with the a$_0$ quoted in Refs. \cite{SolidStateSci.14.1211.2012} and \cite{PhysChemChemPhys.16.345.2014}, although the $c_0$ parameter is closer to our PBEsol results than the PBE values quoted in Ref. \cite{PhysicaB.406.287.2011}, which actually show better agreement with experimental results.\cite{JPhysChemC.120.10547.2016,PhysChemMinerals.30.463.2003} 

Our results obtained for Bi$_2$S$_3$ are also consistent with experimental measurements and other theoretical studies in the literature.
We note that calculations performed on Bi$_2$S$_3$ using the Armiento and Mattsson 2005 parametrized GGA functional (AM05)\cite{JChemPhys.128.084714.2008,PhysRevB.72.085108.2005,PhysRevB.79.155101.2009} seem to show a slightly better reproduction of c$_0$ parameter compared to experiments.\cite{JPhysChemC.120.10547.2016} 

The calculated B$_0$ and B$_0’$ obtained by fitting the $p$-$V$ curves of Sb$_2$Se$_3$ to a third-order Birch-Murnaghan equation are B$_0$ = 31.1 GPa (B$_0'$ = 6.6), which is close to the experimental value of B$_0$ = 30 GPa (B$_0'$ = 6.1) from Ref. \cite{SciRep.3.2665.2013} and  B$_0$ = 32.7 GPa (B$_0'$ = 5.6) from Ref. \cite{CompMatSci.101.301.2015}.
For Sb$_2$S$_3$, we obtained B$_0$ = 31.5 GPa (B$_0'$ = 6.6), which is within the range of experimental values \cite{JPhysChemC.120.10547.2016,PhysChemMinerals.30.463.2003,SciRep.6.24246.2016,SciRep.8.14795.2018} and consistent with the PAW/PBEsol calculations in Ref. \cite{JPhysChemC.120.10547.2016}.
These results are also close to those experimentally measured for the As-doped stibnite mineral.\cite{HighTempHighPress.43.351.2014}
Finally, our values of B$_0$ = 42.3 GPa (B$_0'$ = 6.8) for Bi$_2$S$_3$ are again consistent with other DFT calculations \cite{CompMatSci.101.301.2015,JPhysChemA.118.1713.2014} and the experimental values reported in Refs. \cite{PhysChemMinerals.32.578.2005}, \cite{JPhysChemA.118.1713.2014} and \cite{JAlloysCompd.688.329.2016}.

\subsection{Energetic Stability of the \textit{Pnma} and \textit{R-3m} phases at 0 K and up to 10 GPa}
\label{subsec:thermo}

Since our calculations on the \textit{Pnma} phases were found to be in good agreement with experimental and theoretical studies, we proceeded to carry out a theoretical study of the potential \textit{R-3m} phases of Sb$_2$S$_3$, Bi$_2$S$_3$, and Sb$_2$Se$_3$ to probe whether this phase could be energetically competitive from ambient pressure up to 10 GPa.

\begin{figure}[!]
\begin{center}
\includegraphics[width=8cm]{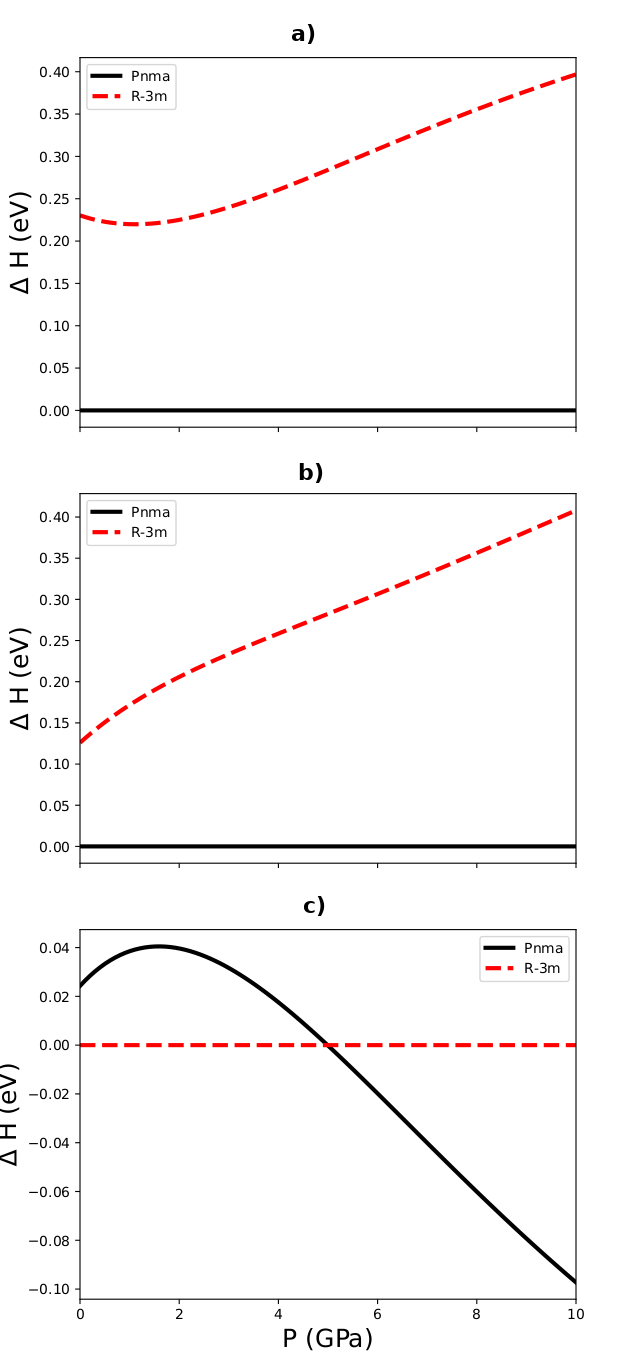}
\caption{\label{fig:enthalpy}
Calculated relative enthalpy \textit{vs} pressure curves up to 10 GPa for  the \textit{Pnma} and \textit{R3m} phases (c.f. Fig \ref{fig:unit-cell}) of Sb$_2$S$_3$ (a), Bi$_2$S$_3$ (b), and Sb$_2$Se$_3$ (c) relative to the predicted lowest-energy phase at ambient pressure, \textit{viz.} the \textit{Pnma} phases of Sb$_2$S$_3$ and Bi$_2$S$_3$ and the \textit{R-3m} phase of Sb$_2$Se$_3$.} 
\end{center}
\end{figure}

Figs. \ref{fig:enthalpy}a, \ref{fig:enthalpy}b, and \ref{fig:enthalpy}c show the pressure-dependence of the enthalpy differences between the \textit{R-3m} and \textit{Pnma} phases of Sb$_2$S$_3$, Bi$_2$S$_3$, and Sb$_2$Se$_3$, respectively.
We find that the orthorhombic \textit{Pnma} phase is the most energetically stable phase of Bi$_2$S$_3$ and Sb$_2$S$_3$ at all pressures examined, as expected from experiments that obtained this phase both under ambient conditions and at high pressure.
Surprisingly, however, our simulations indicate that the \textit{R-3m} phase of Sb$_2$Se$_3$ is more stable than the \textit{Pnma}  below 4.8 GPa, indicating that both the \textit{Pnma} and \textit{R-3m} phases are energetically competitive over this range.
This in principle contradicts existing experimental studies on Sb$_2$Se$_3$ that have so far consistently obtained the \textit{Pnma} phase under ambient conditions.
We note however that the energy difference between the two phases is only 22.71 meV per f.u., which is lower than the k$_\mathrm{B}$T $\sim$ 25 meV at 300 K that would be required for the phase transition to occur under ambient conditions.

In order to rule out the possibility that the above results on the three sesquichalcogenides could be due to inaccuracies with the PBEsol xc functional, we also computed the enthalpies of the two phases with two additional xc functionals, namely the LDA \cite{PhysRevB.23.5048} and the PBE-D2 method using the PBE GGA functional with the Grimme dispersion correction.\cite{JCompChem.27.1787.2006}  
The results of these calculations are presented as an Appendix (Sec. \ref{enthalpy_xc}).
The unit-cell volumes of the \textit{Pnma} phases obtained with the two functionals are underestimated and overestimated with respect to the PBEsol results shown in Tab. \ref{table:pnma_param}, as expected.
As for PBEsol, the two additional functionals predict that at ambient pressure the \textit{R-3m} phase of Sb$_2$Se$_3$ is more energetically stable than the \textit{Pnma} phase, and remains so up to $\sim$6 GPa (LDA) and 8 GPa (PBE-D2).
On the other hand, both functionals predict the \textit{Pnma} phases of Sb$_2$S$_3$ and Bi$_2$S$_3$ are more stable than the \textit{R-3m} phase at least up to 10 GPa.

\subsection{Energetic stability of the low-pressure \textit{Pnma} and \textit{R-3m} phases of Sb$_2$Se$_3$ at finite temperature}
\label{sec:thermo_entropy}

\begin{figure}[!]
\begin{center}
\includegraphics[width=8cm]{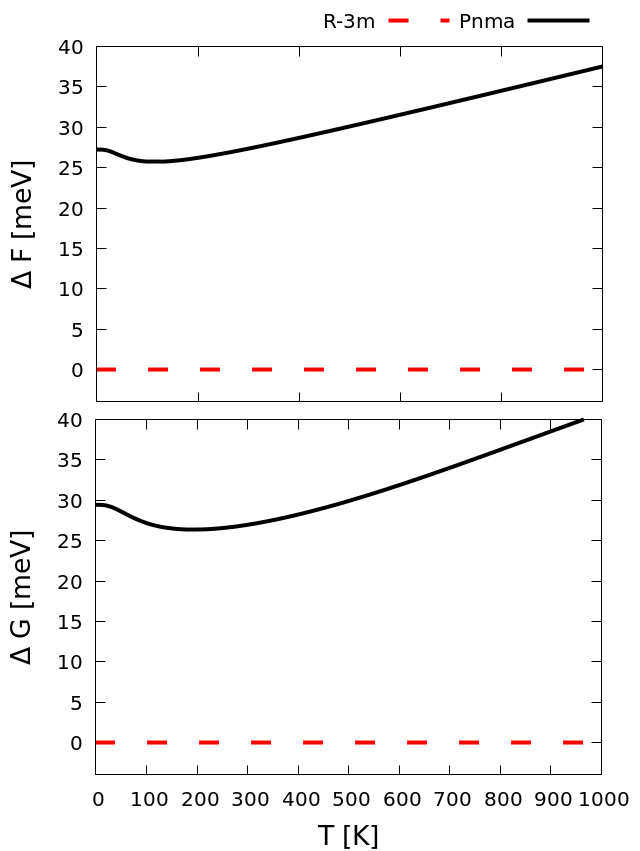}
\caption{\label{fig:free_energies}
Constant-volume Helmoltz (top) and Gibbs (bottom) free energies of the \textit{Pnma} phase (black, solid line) relative to the \textit{R-3m} phase (red, dashed line) of Sb$_2$Se$_3$ as a function of temperature at zero applied pressure.} 
\end{center}
\end{figure}

The calculations in the previous section were performed at 0 K without taking into consideration the contributions to the free energy from crystal vibrations (phonons).
In order to probe whether these effects could alter the energy ordering between the \textit{Pnma} and \textit{R-3m} phases of Sb$_2$Se$_3$, we performed lattice-dynamics calculations on the equilibrium and compressed structures to evaluate the constant-volume Hemholtz and constant-pressure Gibbs free energies at zero pressure ($F$/$G$; Fig. \ref{fig:free_energies}).

The $F$ is obtained by summing the lattice energy (here the DFT total energy) and the vibrational contributions to the internal energy and entropy from the zero-point atomic motion and thermal population of the harmonic phonon energy levels.\cite{JChemPhys.123.204708.2005}
As shown in Fig. \ref{fig:free_energies}, the $F$ predict that at zero pressure the \textit{R-3m} phase remains the most stable across the temperature range examined with no crossing of the free energy to suggest a transition to the \textit{Pnma} phase.
At 0 K, the difference in the $F$ between the two phases is 27.24 meV per f.u., which is $\sim$4.53 meV higher than the difference in $H$ due to the addition of the zero-point energy (i.e. the differences in the phonon frequencies selectively stabilise the \textit{R-3m} phase).
At 300 K, the difference between the phases shows a negligible increase of 0.11 meV per F.U. to 27.35 meV.

Another factor that can influence in the ordering of two competing phases is thermal expansion.
Variation of the lattice volume due to thermal expansion/contraction impacts both the lattice energy and the phonon contributions to the free energy.
This can be accounted for through the quasi-harmonic approximation (QHA) where the thermal expansion of the lattice is predicted from the volume dependence of the lattice energy, phonon frequency spectrum and phonon free energy. \cite{JChemPhys.123.204708.2005, PhysRevB.91.144107.2015}
The free energy $F$ is computed for a series of unit-cell volumes and the equilibrium volume and Gibbs free energy $G$ at a finite temperature $T$ is obtained by minimizing $F$ for a given (constant) pressure.
Fig. \ref{fig:free_energies} shows the difference in $G$ between the \textit{Pnma} and \textit{R-3m} phases of Sb$_2$Se$_3$.
When taking into account the thermal expansion the \textit{R-3m} phase still remains the most stable phase with respect to the \textit{Pnma} phase from 0-1000 K with similar energy difference of 29.43 meV at 0 K to that predicted using the constant-volume $F$, and a slight decrease in the energy difference to 26.96 meV at 300 K.

\begin{figure*}
\begin{center}
\includegraphics[width=14cm]{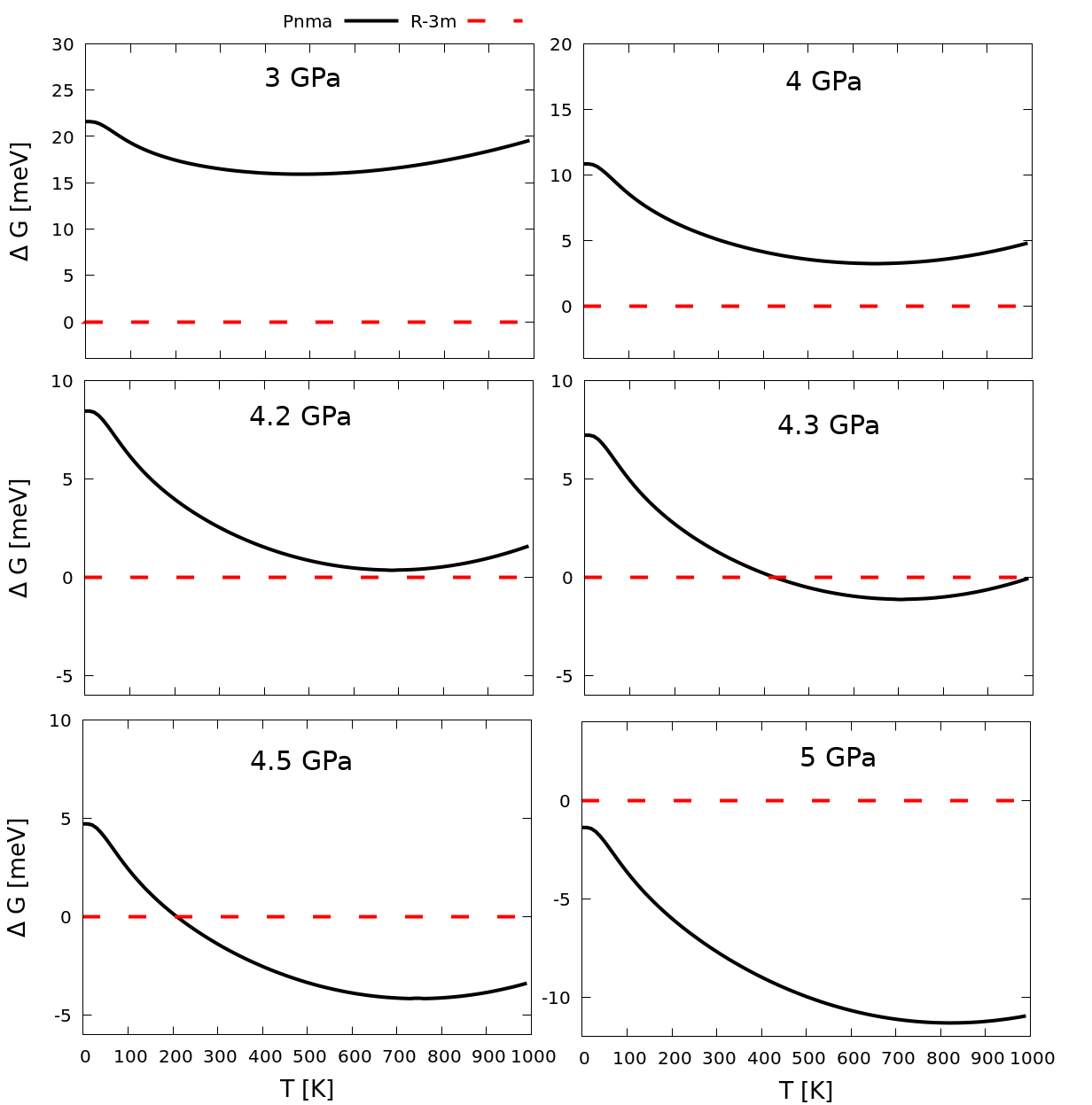}
\caption{\label{fig:free_energies_pressure}
Gibbs free energies of the \textit{Pnma} phase of Sb$_2$Se$_3$ (black, solid line) relative to the \textit{R-3m} phase (red, dashed line)  as a function of temperature at different applied pressures.}
\end{center}
\end{figure*}

To investigate the effect of pressure on the free energy we also computed the difference in the Gibbs energy between the two phases at applied pressures between 0-5 GPa (Fig. \ref{fig:free_energies_pressure}). 
For $p$ = 3 and 4 GPa, these calculations predict the \textit{R-3m} phase to be the most energetically stable phase across the 0-1000 K temperature range examined, but it can be clearly seen that pressure reduces the energy differences between the two phases.
At 4 GPa, the smallest energy difference between the two phases of $\sim$3.25 meV is predicted to occur between 650-700 K.
Increasing the pressure slightly to 4.2 GPa results in the energies of the two phases becoming nearly equal at around 400 K, and at 4.3 GPa a phase transition from \textit{R-3m} to \textit{Pnma} is preducted to occur around this temperature.
At 4.5 GPa the predicted transition temperature decreases to $\sim$200 K, and at 5 GPa the \textit{Pnma} phase becomes the most energetically favorable structure across the entire temperature range examined. 

In summary, free-energy calculations including phonon contributions and thermal expansion therefore at zero applied pressure provude further evidence that the \textit{R-3m} phase of Sb$_2$Se$_3$ is more stable than the \textit{Pnma} phase.
Interestingly, neither the $F$ nor $G$ predict a decrease in the energy difference, from which we infer that phonon contributions to the free energy selectively stabilise the \textit{R-3m} phase, at least at zero applied pressure.
Under low pressures between 4.2-4.4 GPa the Gibbs free energies predict a pressure-induced transition between the \textit{R-3m} and \textit{Pnma} phases near room temperature conditions, which is similar to the 4.8 GPa transition pressure predicted from the 0 K enthalpies without zero-point energy corrections (Fig. \ref{fig:enthalpy}). 
We attribute the small decrease in the predicted transition pressure to the different impacts of volume on the phonon spectra of the two phases.

\subsection{Dynamical stability of the \textit{Pnma} and \textit{R-3m} phases of Sb$_2$Se$_3$}
\label{subsec:phonon}

Energetic stability is a necessary but not sufficient condition for a structural phase to be synthetically accessible.
One should therefore also confirm the dynamical stability of the system, which can be done by studying the phonon frequency spectrum.
If imaginary frequencies are present in the phonon dispersion, this is an indication that the system is not a minimum on the structural potential-energy surface (and is instead e.g. a transition state or hilltop in multidimensional space), and would spontaneously convert to a lower-energy structure and thus be kinetically unstable under a given set of conditions.\cite{ProcCambridgePhilosSoc36.160.1940,Dove.IntLattDyn,Dove.StrutDyn,AmMin.82.213.1997, BullMaterSci.1.129.1979, PhysRevLett.111.025503.2013} 

\begin{figure}
\begin{center}
\includegraphics[width=8cm]{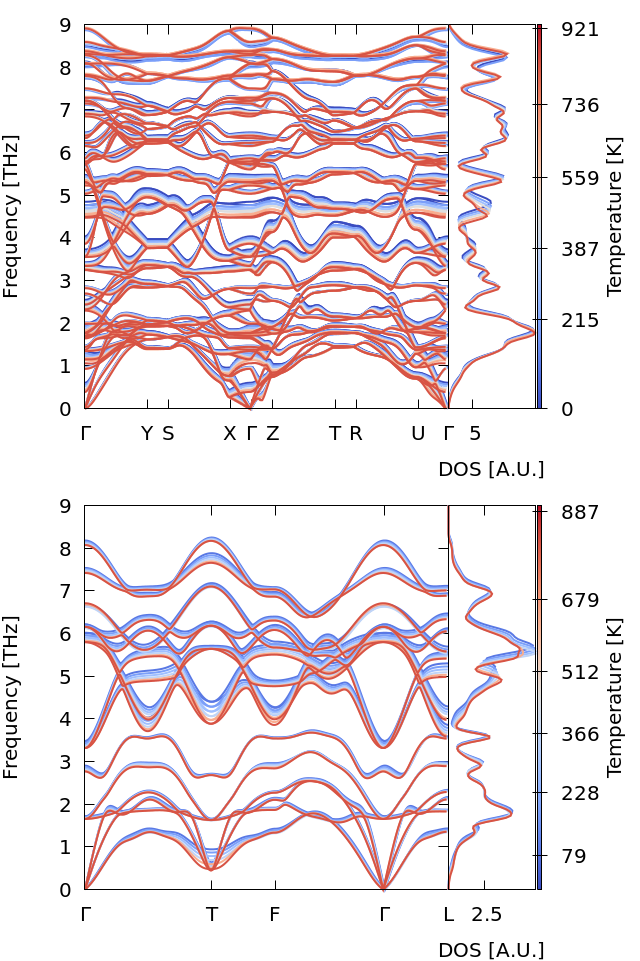}
\caption{\label{fig:qha_band}
Quasi-harmonic phonon dispersion curves for \textit{Pnma}-Sb$_2$Se$_3$ (top) and \textit{R-3m}-Sb$_2$Se$_3$ (bottom). The color gradient runs from blue (low $T$) to red (high $T$) for temperatures between $\sim$0-1000 K. } 
\end{center}
\end{figure}

We therefore investigated dynamical stability of \textit{R-3m} Sb$_2$Se$_3$ to confirm whether this structure could potentially be synthesized at or close to ambient conditions. 
To do so we evaluated the phonon band dispersion and density of states (DoS) curves of the \textit{R-3m} phase at 0 zero pressures and different temperatures using the QHA method described in the previous section (Fig. \ref{fig:qha_band}).
For comparison we also present the phonon band structure and DoS of the \textit{Pnma} phase. Both structures show real frequencies across the whole of the Brillouin zone, indicating that both are dynamically stable under ambient conditions (i.e. 0 GPa and room temperature) and confirming that, as implied by the energetics calculations, both phases should be accessible given appropriate synthesis conditions.

In this context, we note that these results are consistent with a recent theoretical study of the \textit{R-3m} phase of Sb$_2$Se$_3$, which reported phonon dispersion curves and confirmed the dynamical stability of this phase at 0 GPa and 0 K.\cite{PhysRevB.97.075147.2018}
This study also reported formation energies of Sb$_2$Se$_3$ as evidence for the energetic stability of the \textit{R-3m} phase.
In addition, \textit{ab initio} molecular dynamics calculations\cite{PhysRevB.97.075147.2018} have confirmed that the \textit{R-3m} phase remains structurally stable at 300 K.

\subsection{Mechanical stability of the \textit{Pnma} and \textit{R-3m} phases of Sb$_2$Se$_3$}
\label{subsec:elastic}

Having established the energetic and dynamical stability of \textit{R-3m} Sb$_2$Se$_3$ at zero pressure, we also checked the mechanical (elastic) stability as a third condition that must be fulfilled for this system to be synthetically accessible.
This is achieved by calculating the elastic constants and confirming that they obey the Born stability criteria when the solid is subjected to homogeneous deformations.\cite{ProcCambridgePhilosSoc36.160.1940, JPhysCondensMatter.9.8579.1997, PhysRevB.90.224101.2014}
Tab. \ref{table:elastic_cts} lists the calculated elastic constants of the \textit{Pnma} and \textit{R-3m} phases of Sb$_2$Se$_3$ at zero pressure.

\begin{table*}
\begin{center}
\caption{ \label{table:elastic_cts} 
Calculated elastic constants c$_{ij}$ (GPa) of the \textit{Pnma} and \textit{R-3m} phases of Sb$_2$Se$_3$ at zero pressure.}
\begin{tabular}{| c| c|c|c|c|c|c|c|c|c |}\hlinewd{1pt}
  & c$_{11}$  & c$_{22}$ & c$_{33}$ & c$_{12}$ & c$_{13}$ & c$_{23}$ & c$_{44}$ & c$_{55}$ & c$_{66}$\\ \hline
\textbf{\textit{Pnma}} & 30.92  & 81.65 & 55.20 &  17.32 &  15.10 & 26.89 & 17.83 & 25.21 & 7.69 \\
\hlinewd{1pt}
\end{tabular}
\begin{tabular}{| c|c|c|c|c|c|c|c |}\hlinewd{1pt}
& c$_{11}$ = c$_{22}$ & c$_{33}$ & c$_{12}$ & c$_{13}$ = c$_{23}$ & c$_{15}$ = -c$_{25}$ = c$_{46}$  & c$_{44}$ = c$_{55}$ & c$_{66}$\\\hline
\textbf{\textit{R-3m}}  & 90.81  & 40.21 & 25.85 &  21.61 &  -12.00 &  25.10 & 32.48\\ 
\hlinewd{1pt}
\end{tabular}
\end{center}
\end{table*}

To confirm the accuracy of our calculated elastic constants, we computed the linear compressibility of both phases using the ELATE analysis tools. \cite{JPhysCondensMatter.28.275201.2016}
For both phases, only directions corresponding to positive linear compressibilities were obtained, indicating that both phases are mechanically stable under ambient conditions.
In the case of the \textit{R-3m} phase, we obtained linear compressibilities between $\beta_\textrm{min}$ = 4.9 TPa$^{-1}$ (hexagonal $a$-axis) and $\beta_\textrm{max}$ = 19.5 TPa$^{-1}$ (hexagonal $c$-axis) with an anisotropy value of 3.95.
For the \textit{Pnma} phase, the compressibilities fall between $\beta_\textrm{min}$ = 3.7 TPa$^{-1}$ ($b$-axis) and $\beta_\textrm{max}$ = 25.7 TPa$^{-1}$ ($a$-axis) with an anisotropy of 6.87.
These values are of the same order as the experimental axial compressibilities of the \textit{Pnma} phase of Sb$_2$Se$_3$ ($\beta_\textrm{a}$ = 15.2 TPa$^{-1}$, $\beta_\textrm{b}$ = 3.9 TPa$^{-1}$, $\beta_\textrm{c}$ = 8.3 TPa$^{-1}$).\cite{JPhysChemA.118.1713.2014}
Tab. \ref{table:elastic_prop} summarises the elastic moduli calculated from the elastic constants, and we note that the bulk modulus of \textit{Pnma} Sb$_2$Se$_3$ (31.8 GPa) is similar to that obtained from the Birch-Murnaghan fit (c.f. Tab. \ref{table:pnma_param}), which confirms that the elastic constants are adequately converged.

\begin{table}
\begin{center}
\caption{ \label{table:elastic_prop} 
Calculated elastic properties of the \textit{Pnma} and \textit{R-3m} phases of Sb$_2$Se$_3$ at zero pressure obtained within the Voigt approximation using the ELATE analysis tool: bulk modulus, $B_0$, Young modulus, $E$, shear modulus, $G$ and Poisson’s ratio, $\upsilon$.}
\begin{tabular}{| c| c|c|c|c|}\hlinewd{1pt}
  & \textbf{B$_0$ (GPa)} &	\textbf{$E$ (GPa)} &	\textbf{$G$ (GPa)} &  $\mathbf{\upsilon}$\\ \hline
\textbf{\textit{Pnma}} & 31.82	& 44.10	& 17.38 &	0.27 \\ \hline
\textbf{\textit{R-3m}}  & 40.00	& 65.56 &	26.72 &	0.23\\ 
\hlinewd{1pt}
\end{tabular}
\end{center}
\end{table}

The calculated elastic constants in Tab. \ref{table:elastic_cts} fulfill the necessary and sufficient Born criteria for the mechanical stability of orthorhombic (Eq. \ref{eq:pnma}) and rhombohedral (Eq. \ref{eq:r-3m}) systems.\cite{PhysRevB.90.224101.2014}
The calculated elastic constants therefore indicate that both the \textit{R-3m} and \textit{Pnma} phases of Sb$_2$Se$_3$ are mechanically stable under ambient conditions.

\begin{eqnarray}
&& c_{11},c_{44},c_{55},c_{66}>0; \nonumber\\
&& c_{11}c_{22}>c_{12}^2; \nonumber\\ 
&& c_{11}c_{22}c_{33}+2c_{12}c_{13}c_{23}\nonumber\\
&&-c_{11}c_{23}^2-c_{22}c_{13}^2-c_{33}c_{12}^2>0 \label{eq:pnma}
\\\nonumber\\
&& c_{11}>|c_{12}|; c_{44}>0;\nonumber\\
&& c_{13}^2<\frac{1}{2}c_{33}(c_{11}+c_{12}); \nonumber\\
&& c_{14}^2<\frac{1}{2}c_{44}(c_{11}-c_{12})=c_{44}c_{66} 
\label{eq:r-3m}
\end{eqnarray}

The elastic tensors of the \textit{Pnma} phase of Sb$_2$Se$_3$ have been calculated previously calculated.\cite{SolidStateSci.14.1211.2012}
However, in these calculations the components were overestimated as clearly shown by comparison of the bulk modulus with experiments.
Part of the disagreement could be due to the low cut-off energy used in those calculations.

In summary, our calculations on Sb$_2$Se$_3$ demonstrate that the \textit{Pnma} and \textit{R-3m} phases are energetically competitive under ambient conditions and that both phases are dynamically and mechanically stable.
Given that it should in principle be possible to obtain the \textit{R-3m} phase, but it has yet to be reported experimentally, it is possible that the \textit{R-3m} phase is difficult to form due to kinetic reasons, i.e. that the \textit{Pnma} phase is formed faster than the \textit{R-3m} phase under typical synthesis conditions.
We note that the \textit{R-3m} phase did not appear on the pressure/temperature phase diagram prepared by Pfeiffer \textit{et al.},\cite{pfeiffer.PhDthesis.2009} although this study did not attempt to vary the synthesis conditions at close to ambient pressure, which the present calculations suggest would allow this phase to be obtained.
In any case, our calculations raise the possibility that the \textit{R-3m} phase could potentially be prepared under slightly non-equilibrium conditions.
Therefore, to help identify \textit{R-3m} Sb$_2$Se$_3$ in future experiments, we provide in the following section a reference structure and vibrational spectra at zero pressure.

\subsection{Crystal structure and vibrational spectra of \textit{R-3m} Sb$_2$Se$_3$}

Table \ref{table:lattice_r-3m} detailed the predicted equilibrium crystal structure (lattice parameters and atomic positions) of \textit{R-3m} Sb$_2$Se$_3$ obtained using the PBEsol xc functional. (The calculations with PBE+D2 and LDA yield slightly different lattice parameters, \textit{viz.} PBE+D2 - $a_0$=4.02 \AA, $c_0$=28.81 \AA and $V_0$=403.89 \AA$^3$; and LDA -  $a_0$=3.99 \AA, $c_0$=27.58 \AA, and $V_0$=381.61 \AA$^3$.)
The predicted lattice parameters are in good agreement with those reported for other tetradymite-like sesquichalcogenides.\cite{PhysStatusSolidiB.250.669.2013}
The optimised $a_0$ and $c_0$ are slightly smaller than those of Bi$_2$Se$_3$ and much smaller than those of Sb$_2$Te$_3$.\cite{PhysStatusSolidiB.250.669.2013}

\begin{table}
\begin{center}
\caption{ \label{table:lattice_r-3m} 
Predicted equilibrium lattice parameters and atomic positions for the hexagonal unit cell of the \textit{R-3m} phase of Sb$_2$Se$_3$.}
\begin{tabular}{| c| c|c|c|c|}\hlinewd{1pt}
   \textbf{a$_0$ (\AA)}	& \textbf{c$_0$ (\AA)}	& \textbf{V$_0$ (\AA$^3$)} &	\textbf{B$_0$ (GPa)} &	\textbf{B$_0'$}\\ \hline
 4.01 &	28.16 &	392.16	& 50.56	& 4.16 \\ 
\hlinewd{1pt}
\end{tabular}
\begin{tabular}{| c|c|c|c|c|c|}\hlinewd{1pt}
& \textbf{Site}	& \textbf{Sym.}	& \textbf{x}	 & \textbf{y}	& \textbf{z}\\\hline
\textbf{Sb$_1$}	 & 6c	& 3m &	0.00000	& 0.00000	& 0.60082 \\ \hline
\textbf{Se$_1$}	& 3a	 & -3m	& 0.00000	& 0.00000	& 0.00000\\  \hline
\textbf{Se$_2$}	& 6c & 	3m	& 0.00000	& 0.00000	& 0.78792\\ 
\hlinewd{1pt}
\end{tabular}
\end{center}
\end{table}

\begin{figure}
\begin{center}
\includegraphics[width=8cm]{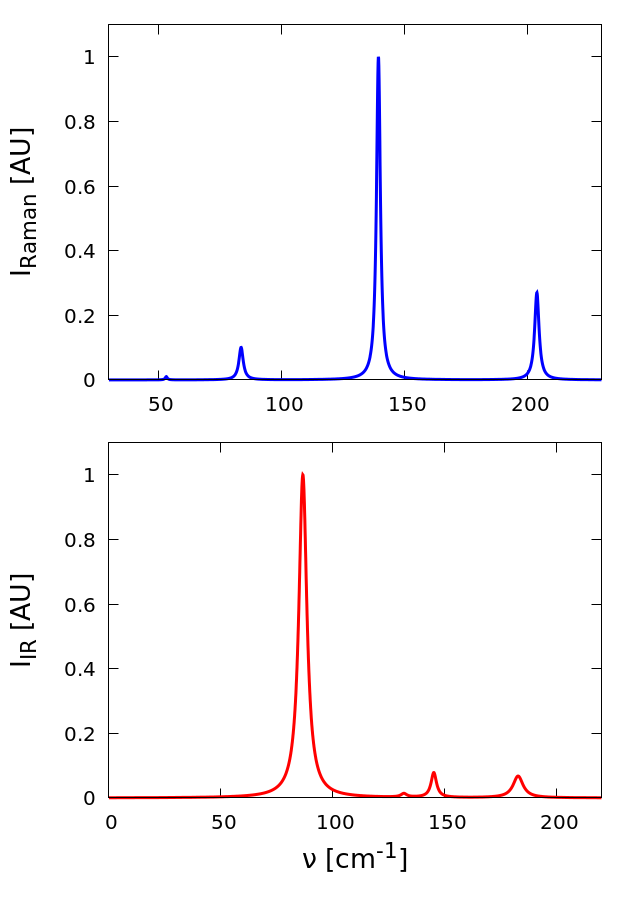}
\caption{\label{fig:raman_Sb2Se3}
Simulated Raman (top) and infrared (IR; bottom) spectra of equilibrium \textit{R-3m} Sb$_2$Se$_3$. The spectral lines have been broadened with the calculated intrinsic mode linewidths at 300 K.} 
\end{center}
\end{figure}

We have also computed the Raman and infrared (IR) spectra of the equilibrium structure to provide spectral signatures that could be used to identify the \textit{R-3m} phase in experiments using routine characterisation techniques (Fig. \ref{fig:raman_Sb2Se3}).
 
The frequencies, irreducible representations and IR/Raman intensities associated with each of the zone-centre ($\Gamma$-point) vibrational modes are listed in Table \ref{table:ir-raman_r-3m}.

\begin{table*}
\begin{center}
\caption{ \label{table:ir-raman_r-3m} 
Calculated vibrational modes of the \textit{R-3m} phase of the equilibrium structure of Sb$_2$Se$_3$. The three acoustic modes span an irreducible representations of $\Gamma_\textrm{acoustic}$ = A$_\textrm{2u}$ + E$_\textrm{u}$, and the remaining 12 optical modes span a representation of $\Gamma_\textrm{optical}$ = 2E$_\textrm{g}$ (Raman) + 2A$_\textrm{1g}$ (Raman) + 2E$_\textrm{u}$ (IR) + 2A$_\textrm{2u}$ (IR).}
\begin{tabular}{| c| c|c|c|c|}\hlinewd{1pt}
 \textbf{ Irr.eps.}	& \textbf{Frequency}	& \textbf{Raman Intensity} 	& \textbf{IR Intensity} &  \textbf{Spectral Linewidths} \\
         	& \textbf{(cm$^{-1}$)}	& \textbf{(10$^5$ \AA$^4$ amu$^{-1}$)}	& \textbf{(e$^2$ amu$^{-1}$)} & \textbf{(cm$^{-1}$)} \\ \hline
 \textbf{E$_\textrm{g}$} &	53.3 & 	0.02 &	Inactive & 1.02\\ \hline
\textbf{A$_\textrm{1g}$} &	83.7	 & 0.37	& Inactive & 2.11\\ \hline
\textbf{E$_\textrm{u}$} & 	86.6	 & Inactive	& 3.47 &  4.32\\ \hline
\textbf{E$_\textrm{u}$}& 	131.7 &	Inactive &	0.03 & 3.13 \\  \hline
\textbf{E$_\textrm{g}$} & 	139.4 & 3.32	 & Inactive & 1.89  \\ \hline
\textbf{A$_\textrm{2u}$} &	145.1	& Inactive	& 0.18 &   2.94 \\ \hline
\textbf{A$_\textrm{2u}$} &	182.8	& Inactive & 	0.29 &  5.38 \\ \hline
\textbf{A$_\textrm{1g}$} &	203.8	& 1.02	& Inactive & 2.16\\
\hlinewd{1pt}
\end{tabular}
\end{center}
\end{table*} 
 
The inversion symmetry in the \textit{R-3m} structure leads to mutual exclusion between the IR and Raman activity of the modes, with each spectrum being characterized by four bands.\cite{PhysStatusSolidiB.250.669.2013}
The most intense Raman band occurs around 139 cm$^{-1}$ (E$_\textrm{g}$), while a second prominent feature is predicted at $\sim$204 cm$^{-1}$ (A$_\textrm{1g}$).
The frequency of this A$_\textrm{1g}$ mode is higher than in Bi$_2$Se$_3$ but lower than in In$_2$Se$_3$, as expected from the difference in mass between In, Sb and Bi.\cite{PhysRevB.84.184110.2011,InorgChem.57.8241.2018}
Lower-frequency E$_\textrm{g}$ and A$_\textrm{1g}$ modes with much lower intensities are also found around 53 and 84 cm$^{-1}$, respectively, which are again slightly blue-shifted when compared to the corresponding frequencies calculated for Bi$_2$Se$_3$.\cite{PhysRevB.84.184110.2011}
There are four IR-active modes, two with E$_\textrm{1u}$ symmetry (87 and 132 cm$^{-1}$) and two with A$_\textrm{2u}$ symmetry (145 and 183 cm$^{-1}$).
Of these, the 87 cm$^{-1}$ mode is the most prominent in the spectrum, while the second E$_\textrm{u}$ mode at 132 cm$^{-1}$ is comparatively weak.
The two A$_\textrm{2u}$ bands have moderate ans comparable intensities and form a pair of smaller features at higher frequencies.
As expected given the mass difference, the IR-active modes in Sb$_2$Se$_3$ again have slightly higher frequencies than those calculated for Bi$_2$Se$_3$.\cite{PhysRevB.84.184110.2011}

It is also worth comparing our predicted structural and vibrational properties of \textit{R-3m} Sb$_2$Se$_3$ to the published works pertaining to the possible synthesis of this phase.
In 2013, Bera \textit{et al.} \cite{PhysRevLett.110.107401.2013} claimed to observed the \textit{R-3m} phase of Sb$_2$Se$_3$ at room temperature. 
However, the experimental lattice parameters were not disclosed and the reported Raman spectrum is not consistent with our theoretical spectrum, although our spectrum is consistent with simulations performed in the same study.
We also note that our calculations of the pressure coefficients for the Raman- and IR-active modes of tetradymite Sb$_2$Se$_3$ (not shown) indicate that all the modes should show positive pressure coefficients, which is again consistent with the theoretical results in Ref. \cite{PhysRevLett.110.107401.2013}.

If we consider that the band gap of tetradymite-like Sb$_2$Se$_3$ should be similar to or even larger than that of tetradymite-like Bi$_2$Se$_3$, we would expect the Raman spectrum of tetradymite-like Sb$_2$Se$_3$ to be similar.
We may therefore conclude that the experimental Raman spectra reported in Ref. \cite{PhysRevLett.110.107401.2013} is not consistent with tetradymite-like Sb$_2$Se$_3$.
In this vein, we note that the appearance of soft Raman modes with negative pressure coefficient at low pressure as reported in Ref. \cite{PhysRevLett.110.107401.2013} are similar to those of Se and Te nano- or micro-clusters.\cite{doi:10.1021/jp402493r,doi:10.1021/acs.jpcc.6b06049,PhysRevB.94.134102,D1TC00980J} Such features are either formed during synthesis or induced by laser heating in Raman scattering measurements at high laser powers, as has recently been discussed.\cite{D1TC00980J}   

Recently, Matetskiy et al. \cite{Matetskiy_2020} claim to have observed the tetradymite-like phase of Sb$_2$Se$_3$ in MBE-deposited layers over a thick buffer layer of tetradymite-like Bi$_2$Se$_3$.
This study reported a lattice parameter of $a_0$=4.048~\AA~ for 5 quintuple layers of tetradymite-like Sb$_2$Se$_3$.
The quintuple layers were observed in scanning-tunnelling microscopy (STM) to be of $\sim$1 nm in thickness, meaning that the $c_0$ lattice parameter, corresponding to three quintuple layers, would be $\sim$30~\AA.
Both the reported $a_0$ and $c_0$ lattice parameters are consistent with our calculated values.
Furthermore, electron dispersion obtained from angle-resolved photoemission spectroscopy (ARPES) is compatible with the \textit{R-3m} phase.
It is therefore highly probable that tetradymite-like Sb$_2$Se$_3$ was synthesized for the first time in this study.
Unfortunately, this study did not report vibrational spectra that would allow for additional comparison to our predictions.

\subsection{On the observation of the \textit{R-3m} phase in \textit{A$_2$X$_3$}  sesquichalcogenides}

An analysis of the bonding in layered materials under ambient conditions, including the tetradymite-like B$^V_2$X$^{VI}_3$ and A$^{VI}$B$^V_2$X$^{VI}_4$ chalcogenides and also the transition metal dichalcogenides (TMDs), has recently been performed by plotting the van der Waals (vdW) gap spacing \textit{vs.} the X-X plane spacing.\cite{adfm.201705901}
The vdW gap spacing is defined as the interplanar distance across the interlayer space, whereas the X-X plane distance is the interplanar width across the intralayer space.\cite{00018736900101307, adfm.201705901}
In Ref. \cite{adfm.201705901}, it was concluded from the strong correlation that TMDs show a pure interlayer vdW interaction, whereas the tetradymite-like chalcogenides do not because the vdW gap spacing is much smaller than expected for their X-X plane spacing, which is indicative of a stronger interlayer interaction than that expected for pure vdW interactions.

It has recently been suggested that the vdW gap spacings in tetradymite-like group-15 sesquichalcogenides are smaller than those found for pure vdW materials, e.g. GaSe, InSe and
TMDs. This has been attributed to the presence of the extra delocalized electrons between the layers that contribute an electrostatic component to the bonding that is not present in pure
vdW materials.\cite{adma.201904316,adma.201908302} These delocalized electrons arise from a new type of bonding, termed "metavalent bonding", between the cations and anions inside the quintuple layers  of the tetradymite-like structure, which has been demonstrated by the observation of a net charge difference between the layers in group-15 sesquichalcogenides \cite{JApplPhys.121.225902.2017} and through a topological study of the electronic charge density of SnSb$_2$Te$_4$.\cite{acs.inorgchem.0c01086}
In essence, metavalent bonding results in the cations and anions providing extra electrons to the space between the layers, which in turn supports a stronger interlayer interaction. 

\begin{figure}
\begin{center}
\includegraphics[width=9cm]{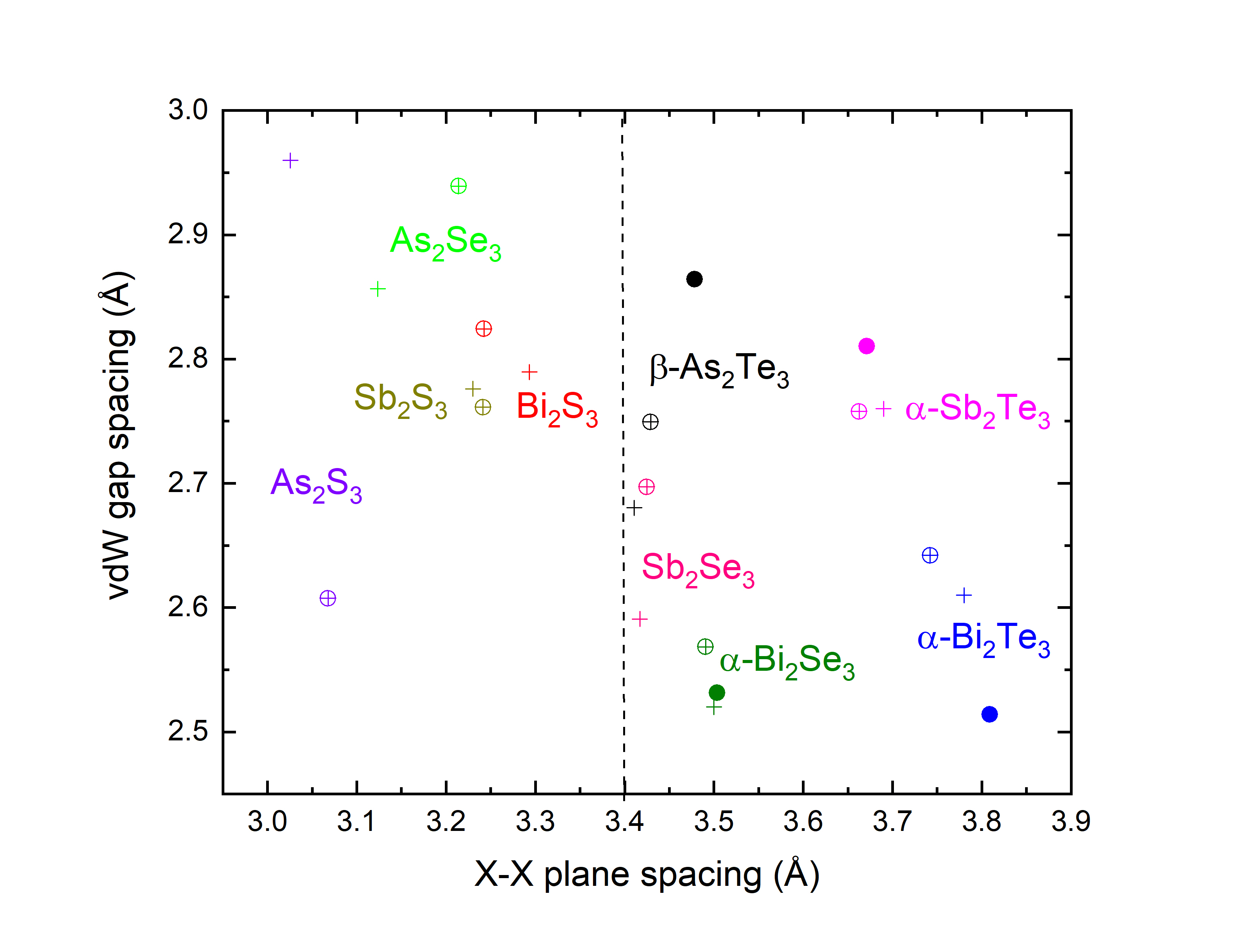}
\caption{\label{fig:meta}
Size of van der Waals gap \textit{vs.} chalcogenide X–X plane spacing in the \textit{R-3m} phases of the \textit{A$_2$X$_3$} sesquichalcogenides (\textit{A} = As, Sb, Bi and \textit{X} = Te, Se, S).
The data obtained from our theoretical calculations using dispersion-corrected PBE are shown by circles with crosses.
For comparison we also show theoretical PBEsol calculations on $\alpha$-Bi$_2$Se$_3$, $\alpha$-Bi$_2$Te$_3$ and $\alpha$-Sb$_2$Te$_3$ from Ref. \cite{00018736900101307} (crosses) and experimental measurements on $\beta$-As$_2$Te$_3$, $\alpha$-Bi$_2$Se$_3$, $\alpha$-Bi$_2$Te$_3$ and $\alpha$-Sb$_2$Te$_3$ from Refs. \cite{doi:10.1021/acs.inorgchem.5b01676,NAKAJIMA1963479,ADAM20071986,Anderson:a11041} (dots).} 
\end{center}
\end{figure}

In Fig. \ref{fig:meta} we compare the experimental and theoretical vdW gap spacing \textit{vs.} the X-X plane spacing for all tetradymite-like B$^V_2$X$^{VI}_3$ (B=As,Sb,Bi; X=S,Se,Te) sesquichalcogenides. \cite{00018736900101307, adfm.201705901} 
(The theoretical results are obtained with PBEsol and dispersion-corrected PBE.)

Fig. \ref{fig:meta} shows that the compounds usually synthesized in the tetradymite-like structure (i.e. $\beta$-As$_2$Te$_3$, $\alpha$-Sb$_2$Te$_3$, $\alpha$-Bi$_2$Te$_3$ and $\alpha$-Bi$_2$Se$_3$) have a very small vdW gap spacing (typically below 2.9~\AA) and a large X-X plane spacing (typically above 3.4~\AA).
It is rather striking that the Te-Te plane spacing in $\beta$-As$_2$Te$_3$ is below 3.5~\AA, which is quite far from the values for other Te sesquichalcogenides and close to the values of the Se-Se plane spacings in Se sesquichalcogenides such as Bi$_2$Se$_3$.
In this regard, it can be inferred that all compounds with a stable tetradymite-like structure at room conditions (see Fig. 5 of Ref. \cite{adfm.201705901}) show vdW-gap spacings distributed in a narrow X-X plane spacing region between 3.4 and 3.8~\AA.
Since As$_2$Se$_3$ and Sb$_2$Se$_3$ do not usually crystallize in the tetradymite-like structure, it could be similarly inferred that the region located between 3.4~\AA < X-X plane spacing < 3.8~\AA~and a vdW gap spacing < 2.9~\AA~imposes a limit of stability for the formation of the tetradymite structure in B$^V_2$X$^{VI}_3$ and A$^{VI}$B$^V_2$X$^{VI}_4$ chalcogenides.
This is consistent with: i) the location of $\alpha$-Bi$_2$Se$_3$ and $\beta$-As$_2$Te$_3$ close to the lower limit value of X-X plane spacing; ii) the metastable tetradymite-like phase of $\beta$-As$_2$Te$_3$ obtained under ambient conditions; and (iii) the metastable orthorhombic \textit{Pnma} phase of Bi$_2$Se$_3$ (guanajuatite) under ambient conditions,\cite{FizTverdTela.6.2223.1964,KANG2017223,FizTverdTela.75.3508.1973, Kristall.18.173.1973} which is isostructural to Bi$_2$S$_3$, Sb$_2$S$_3$ and Sb$_2$Se$_3$ and has been found in HP studies on $\alpha$-Bi$_2$Se$_3$ (paraguanajuatite) on decreasing the pressure.\cite{PhysChemChemPhys.16.345.2014,HighTempHighPress.43.351.2014}

Most importantly, the vdW gap spacings and X-X plane spacings calculated for Sb$_2$Se$_3$ with PBEsol and PBE, with dispersion corrections, are similar to those obtained for group-15 tetradymite-like sesquichalcogenides. This lends further support to the idea that the \textit{R-3m} phase of Sb$_2$Se$_3$ should be synthetically accessible with a structure similar to that predicted in our calculations. Unfortunately, despite the fact that \textit{R-3m} Sb$_2$Se$_3$ does appear to have been synthesized recently,\cite{Matetskiy_2020, adfm.201705901} a complete structural characterization has yet to be reported and therefore several of our calculated parameters cannot be compared to experimental.
We thus hope that the present study may stimulate further attempts to prepare and characterise the \textit{R-3m} phase of Sb$_2$Se$_3$. 

Finally, by observing Fig. \ref{fig:meta} we also note that all group-15 sesquisulphides, for which theoretical \textit{R-3m} structures are located below 3.4~\AA~of the X-X gap spacing, do not seem to be stable in this phase at room conditions according to our calculations.
This is in good agreement with the high predicted enthalpies of this phase of Sb$_2$S$_3$ and Bi$_2$S$_3$.
However, this does not necessarily mean that the \textit{R-3m} phase of these compounds could not be synthesized, as it is known that \textit{R-3m} As$_2$Se$_3$ can be prepared at high pressure and high temperature, which indicates that the tetradymite-like structure of As$_2$Se$_3$ is metastable under ambient conditions.\cite{pfeiffer.PhDthesis.2009,LITYAGINA2015799,Novosibirsk.1985,FizTverdTela.11.2382.1969}
It is therefore possible that tetradymite-like As$_2$S$_3$, Sb$_2$S$_3$ and Bi$_2$S$_3$ could also potentially be synthesized.

\section{Conclusions}

We have performed a comparative theoretical study of the \textit{Pnma} and \textit{R-3m} phases of Sb$_2$S$_3$, Bi$_2$S$_3$, and Sb$_2$Se$_3$ considering applied pressures of up to 10 GPa.
Our calculations predict that at ambient pressure the \textit{R-3m} (tetradymite-like) phase of Sb$_2$Se$_3$ is energetically more stable than the \textit{Pnma} phase, in contrast to Sb$_2$S$_3$ and Bi$_2$S$_3$.
This result contradicts the fact that all three compounds are usually grown on the \textit{Pnma} phase.
Further energetic studies of both Sb$_2$Se$_3$ phases show that the higher energetic stability of the \textit{R-3m} phase with respect to the \textit{Pnma} phase is predicted by three different xc functionals and is unaffected by the phonon contributions to the free energy and thermal expansion at finite temperature.
Lattice dynamics and elastic tensor calculations further show that both phases of Sb$_2$Se$_3$ are dynamically and mechanically stable at zero pressure
Our calculations therefore suggest that the formation of this phase should be feasible at close to ambient conditions.

To aid in its identification, we have provided a theoretical crystal structure and predicted IR and Raman spectra.
We also discussed the results of the only two published works, to the best of our knowledge, that have claimed to have synthesized tetradymite-like Sb$_2$Se$_3$, and concluded that there is a high probability that this phase has recently been synthesized by molecular beam epitaxy on a thick buffer layer of tetradymite-like Bi$_2$Se$_3$.

Finally, we have discussed the stability and metastability of the \textit{R-3m} structure for all group-15 sesquichalcogenides on the basis of the values of the vdW gap spacings and the X-X plane spacings, which again suggests that the \textit{R-3m} phase of Sb$_2$Se$_3$ should be synthetically accessible.
We hope that this work will stimulate further investigation of tetradymite-like Sb$_2$Se$_3$ as well as the corresponding phase of As$_2$Se$_3$ and other sesquichalcogenides. These type of sesquichalcogenides could potentially show topological properties interesting for spintronics and quantum computation at moderate pressures, and also have interest as phase change materials and highly-efficient thermoelectric materials, as well as for photonic devices.

\section*{Author Contributions}
E.L.d.S., M.C.S., J.M.S., P.R.H. and A.M. performed the \textit{ab-initio} calculations. 
E.L.d.S., M.C.S., J.M.S., P.R.H., A.M., D.M.G., R.V., and F.J.M. contributed to the interpretation and discussion of results.
F.J.M. supervised the project. 
E.L.d.S., M.C.S., J.M.S., P.R.H., A.M., D.M.G., R.V., and F.J.M. have contributed equally to the discussion and drafting of the paper.
All authors have read and agreed to the published version of the manuscript.

\section*{Conflicts of interest}
There are no conflicts to declare.

\section*{Acknowledgements}
This publication is part of the project MALTA Consolider Team network (RED2018-102612-T), financed by MINECO/AEI/10.13039/501100003329; by I+D+i projects PID2019-106383GB41/42/43, financed by MCIN/AEI/10.13039/501100011033; by project PROMETEO/2018/123 (EFIMAT), financed by Generalitat Valenciana; and by the European Union Horizon 2020 research and innovation programme under Marie Sklodowska-Curie grant agreement No. 785789-COMEX. E.L.d.S., A. M., and P. R.-H. acknowledge computing time provided by MALTA-Cluster at the University of Oviedo and Red Española de Supercomputación (RES) through the computer resources at MareNostrum with technical support provided by the Barcelona Supercomputing Center (QCM-2018-3-0032). E.L.d.S. also acknowledges the Network of Extreme Conditions Laboratories (NECL), financed by FCT and co-financed by NORTE 2020, through the programme Portugal 2020 and FEDER. J.M.S. is grateful to UK Research and Innovation for the support of a Future Leaders Fellowship (MR/T043121/1) and to the the University of Manchester for the previous support of a Presidential
Fellowship.

\section*{Data availability}

Raw data from this study can be obtained from the corresponding author on reasonable request.

\section*{License statement}
For the purpose of open access, the author has applied a Creative Commons Attribution (CC BY) licence to any Author Accepted Manuscript version arising.

\clearpage
\section{Appendix}
%

\subsection{Enthalpy \textit{vs} pressure curves obtained using different exchange-correlation Functionals}\label{enthalpy_xc}

\begin{figure}
\begin{center}
\includegraphics[width=8cm]{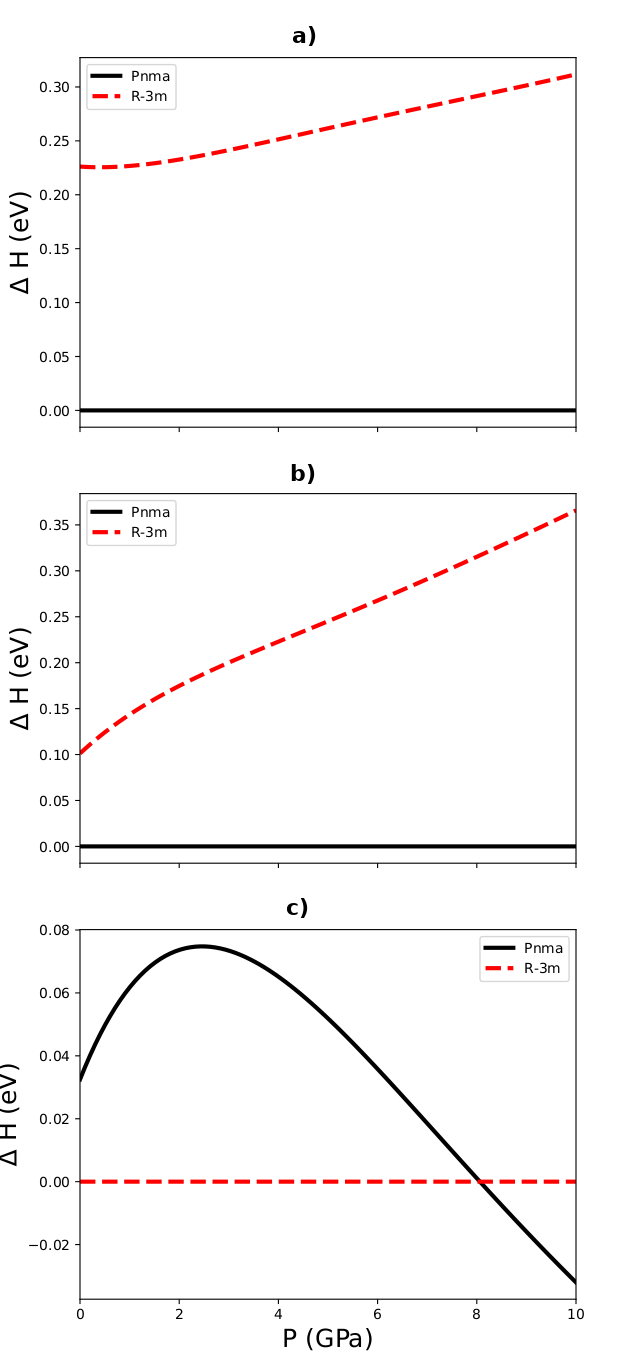}
\caption{\label{fig:enthalpy_pbevdw}
PBE-D2 enthalpy \textit{vs} pressure curves for the \textit{Pnma} and \textit{R-3m} phases of Sb$_2$S$_3$(a), Bi$_2$S$_3$(b)and Sb$_2$Se$_3$(c), relative to the lowest-energy phase at ambient pressure, \textit{viz.} the \textit{Pnma} phases of Sb$_2$S$_3$ and Bi$_2$S$_3$ and the \textit{R-3m} phase of Sb$_2$Se$_3$.} 
\end{center}
\end{figure}

\begin{figure}
\begin{center}
\includegraphics[width=8cm]{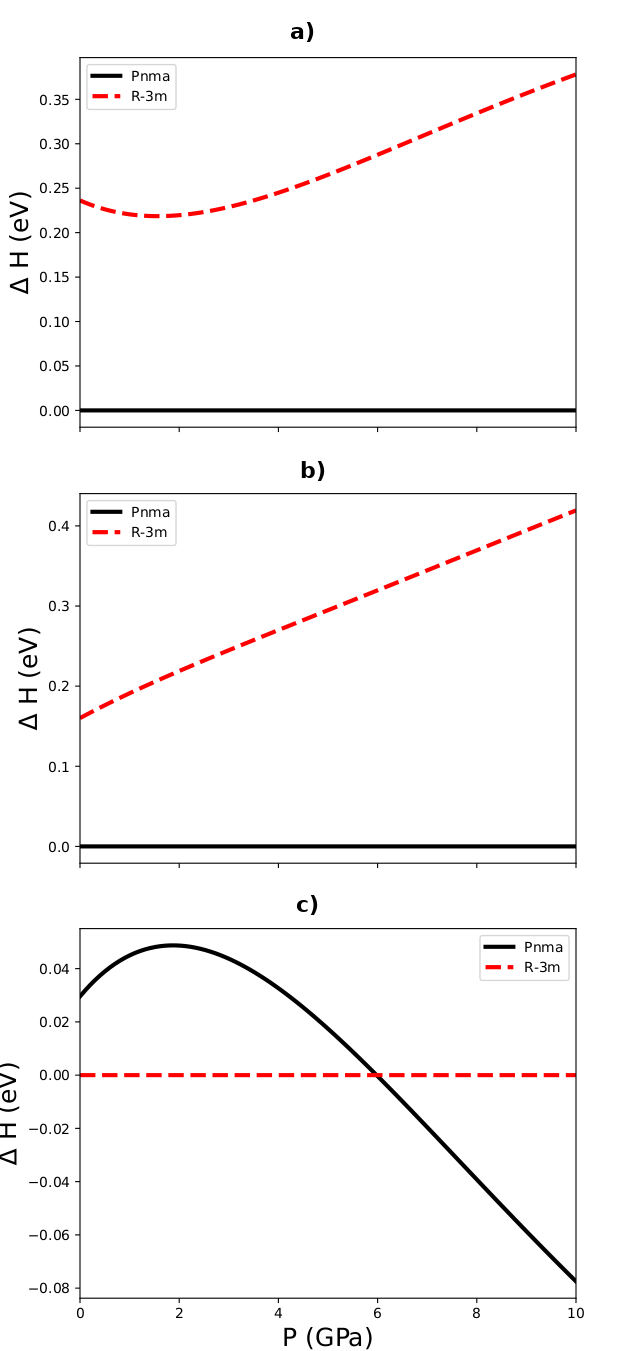}
\caption{\label{fig:enthalpy_lda}
LDA enthalpy \textit{vs} pressure curves for the \textit{Pnma} and \textit{R-3m} phases of Sb$_2$S$_3$(a), Bi$_2$S$_3$(b)and Sb$_2$Se$_3$(c), relative to the lowest-energy phase at ambient pressure, \textit{viz.} the \textit{Pnma} phases of Sb$_2$S$_3$ and Bi$_2$S$_3$ and the \textit{R-3m} phase of Sb$_2$Se$_3$.} 
\end{center}
\end{figure}

In order to confirm the variation of the enthalpy with pressure \textit{Pnma} and \textit{R-3m} phases of the three sesquichalcogenides presented in Fig. \ref{fig:enthalpy} and obtained using the PBEsol functional, we performed similar calculations using dispersion-corrected PBE (PBE-D2; Fig. \ref{fig:enthalpy_pbevdw}) and the LDA (\ref{fig:enthalpy_lda}).
For the Sb$_2$Se$_3$ we find that both additional functionals also predict that the \textit{R-3m} phase is energetically the most stable under ambient conditions.
There is however some variation in the predicted transition pressure, with PBEsol and the LDA predicting similar pressures but PBE-D2 predicting a slightly higher transition point.

\balance

\clearpage
\providecommand*{\mcitethebibliography}{\thebibliography}
\csname @ifundefined\endcsname{endmcitethebibliography}
{\let\endmcitethebibliography\endthebibliography}{}

\end{document}